# $Cs_3V_9Te_{13}$: A Correlated Electron System with Topological Flat Bands

Chang-Chao Liu,[1*] Ji-Yong Liu,[2*] Jing Li,[1*] Hua-Xun Li,[2,3] Jia-Yi Lu,[1] Tong Shi,[4] Qing-Xin Dong,[4] Gen Li,[4] Bo-Sen Wang,[4] Yi Liu,[5†] Jin-Guang Cheng,[4] and Guang-Han Cao[1,7 †]

[1]School of Physics, Zhejiang University, Hangzhou 310058, China

[2]Department of Chemistry, Zhejiang University, Hangzhou 310058, China

[3]School of Physics and Hangzhou Key Laboratory of Quantum Matters, Hangzhou Normal University, Hangzhou 311121, China

[4]Beijing National Laboratory for Condensed Matter Physics and Institute of Physics, Chinese Academy of Sciences, Beijing 100190, China

[5]School of Physics, Key Laboratory of Quantum Precision Measurement of Zhejiang Province, Zhejiang University of Technology, Hangzhou 310023, China

[6]School of Physical Sciences, University of Chinese Academy of Sciences, Beijing 100190, China

[7]Interdisciplinary Center for Quantum Information, and State Key Laboratory of Silicon and Advanced Semiconductor Materials, Zhejiang University, Hangzhou 310058, China

[*]These authors contributed equally

[†]E-mail: liuyiphy@zjut.edu.cn; ghcao@zju.edu.cn

**Abstract**

Correlated electron systems with topological flat bands show great promise in exploring exotic quantum phenomena. However, such crystalline materials remain rare. Here we report the discovery of a novel material, $Cs_3V_9Te_{13}$, which unexpectedly exhibits magnetism and significant electron correlations. The crystal structure features two interpenetrating sets of vanadium triangles that can be linked with an ideal kagome lattice. The physical property measurements demonstrate a cascade of correlated electron phenomena, including quasi-two-dimensional bad metal, non-Fermi-liquid behavior, antiferromagnetic spin-density-wave transition at $T_N$ = 47 K, possible short-range spin ordering at ~350 K, a large Sommerfeld coefficient of 246 mJ mol-fu$^{-1}$ K$^{-2}$, and pressure-induced quantum criticality. These correlated electron behaviors are associated with the topological flat bands at the Fermi level, the latter of which are generated from the V2 sublattice in terms of a bipartite kagome model. Our findings establish Cs$_3$V$_9$Te$_{13}$ as a brand new correlated matter that synergistically combines flat-band physics and tunable properties.

## Introduction

Materials with electronic flat bands (FBs), characterized by a vanishing dispersion near the Fermi level ($E_F$), have emerged as a fertile platform for exploring exotic quantum phenomena.[1] In such systems, the quenched kinetic energy, induced by frustrated lattices, moiré heterostructures, and orbital localizations, enhances the role of Coulomb interactions, giving rise to novel magnetism, unconventional superconductivity, and topologically nontrivial states.[1–4] Currently, the realization of topological FBs in crystalline solids is being pursued in order to investigate strong correlation effects beyond conventional paradigms.

Kagome materials may inherently host topological FBs in addition to the characteristic Dirac cones and van Hove singularities (vHs).[2,5,6] Such FBs arise from the destructive interference of wave functions in the frustrated kagome lattice. While the geometry driven FBs often exist in kagome-structured materials, unfortunately, these FBs are mostly located far away from $E_F$. In this context, the recently discovered kagome metal CsCr$_3$Sb$_5$[4] turns out to be exceptional. With FBs near $E_F$, CsCr$_3$Sb$_5$ exhibits strong electron correlations,[4,7–10] spin- and charge- density waves,[4,11–15] and pressure-induced superconductivity.[4,16] Note that such kagome-like band structures can also be found in a broader class of materials with-out possessing an ideal kagome network.[17] This provides additional possibilities to realize topological FBs in crystalline solids.

In this work, we report on the preparation, crystal structure, physical properties, pressure effect, and electronic structure calculations of a novel vanadium-based material, Cs$_3$V$_9$Te$_{13}$, which was discovered by serendipity while growing crystals of the altermagnet candidate Cs$_{1-\delta}$V$_2$Te$_2$O.[18] The new compound crystallizes in a complex variant of the kagome metals, $A$V$_3$Sb$_5$ ($A$ = K, Rb, Cs), the latter of which have been intensely studied in recent years.[19–25] Nonetheless, the kagome network is absent in Cs$_3$V$_9$Te$_{13}$, and there are two interpenetrating sets of vanadium triangles in the V$_9$Te$_3$ plane. Interestingly, the V2 triangles contribute two sets of kagome-like bands, which by coincidence aligns with the diatomic (bipartite) kagome model theoretically proposed earlier.[26] Importantly, $E_F$ locates just at one of the topological FBs, allowing appearance of magnetism and strong electron correlations. Indeed, we observed a cascade of correlated electron phenomena: bad-metal and non-Fermi-liquid (NFL) behaviors, antiferromagnetic (AFM) spin-density-wave transition, and enhanced Sommerfeld coefficient. The strength of electron correlations can be easily tuned by applying pressure, which shows an unusual quantum criticality associated with two vanadium sub-lattices. Although no superconductivity was observed down to 2 K and up to 9 GPa, the material definitely displays intriguing electron correlation effects due to the topological FB.

## Results and Discussion

### Crystal Structure

$Cs_3V_9Te_{13}$ crystallizes in a hexagonal structure with the space group of $P\bar{6}2m$ (No. 189). As shown in Figure 1, it has a quasi-two-dimensional (Q2D) layered structure containing $[V_9Te_{13}]^-$ slabs stacking with $Cs^+$ counter-ion layers alternatively along the $c$ axis, similar to the architecture in $CsV_3Sb_5$.[27] All the vanadium ions are located in the middle of the $V_9Te_{13}$ slabs, forming the $V_9Te_3$ planes that are sandwiched by the six Te2 and four Te3 atoms. The $a$ and $b$ axes are about √3 times of those of $CsV_3Sb_5$, and the $V_9Te_3$ plane can be linked with the $V_3Sb$ plane in $CsV_3Sb_5$. As shown in Figure 1(d), first of all, the V sublattice in $CsV_3Sb_5$ forms an ideal kagome net, a 2D network of corner-sharing triangles, with an antimony atom at the center of the kagome hexagons. If one of the three V triangles rotates 60°, a 2D √3 ×√3 superlattice will then be resulted. In the real structure of $Cs_3V_9Te_{13}$, the atomic positions of V1, V2, and Te1 are slightly adjusted to minimize the internal energy without altering the structural topology. Correspondingly, Te2 and Te3 change their positions such that V2 is octahedrally coordinated by two Te1, Te2, and Te3 atoms, similar to the case in $CsV_3Sb_5$. Differently, V1 has a tetragonal pyramidal coordination with one Te1 and four Te2 atoms. As a result, two Te atoms are missing, and the expanded unit cell gives the chemical formula $Cs_3V_9Te_{13}$, which is confirmed by the chemical composition measurement (Figure S1). Notably, the interatomic distances in the V1 and V2 triangles are 2.708(6) Å and 3.206(4) Å, respectively. The former is even smaller than that (2.747 Å) in $CsV_3Sb_5$,[27] suggesting stronger V1-V1 interactions. Contrastingly, the V2-V2 distance is remarkably large, implying narrow 3$d$ electronic bands (Figure S4) that could give magnetism and electron correlations.

## Physical Properties

The physical properties of $Cs_3V_9Te_{13}$ crystals are investigated via the magnetic, electrical transport, and thermodynamic measurements. Figure 2(a) shows the temperature depen-dence of anisotropic susceptibility with magnetic fields parallel to the *ab* plane ($\chi_{ab}$) and *c* axis ($\chi_c$). Both $\chi_{ab}(T)$ and $\chi_c(T)$ exhibit a kink at $T_N$ = 47 K, suggesting an AFM transition. Note that $\chi_{ab}$ is more prominently suppressed below $T_N$, implying that the magnetic moments lie within the *ab* plane. Above room temperature, a broad hump shows up at around 350 K, suggesting existence of short-range magnetic order. This susceptibility anomaly is tentatively attributed to V1 triangles, since the exchange interactions are expected to be much larger due to its shorter interatomic distance (see above). In the intermediate temperature range, $\chi(T)$ obeys the extended Curie-Weiss law, $\chi(T) = \chi_0 + C/(T - \theta_W)$. The data fittings in different temperature ranges yield basically consistent fitted parameters (Table S2). As for the data fitting from 50 to 140 K, the estimated effective moments are 0.87 and 0.78 $\mu_B$/V2 for $H \mathbin{/\mkern-6mu/} ab$ and $H \mathbin{/\mkern-6mu/} c$, respectively, assuming that the magnetic moments solely come from the V2 sublattice. These small values of the effective moment suggest itinerant magnetism in $Cs_3V_9Te_{13}$, and thus the AFM order is actually a spin-density wave, like the case in the parent compounds of iron-based superconductors.[28] The fitted paramagnetic Weiss temperatures are $\theta_W$ = −79.2 and −48.8 K, respectively, for $H \mathbin{/\mkern-6mu/} ab$ and $H \mathbin{/\mkern-6mu/} c$. The negative values of $\theta_W$ indicate dominant AFM interactions. Furthermore, there exist in-plane magnetic frustrations in the V2 sublattice, as is reflected by the frustration index, $\theta_W/T_N \approx 1.7$, for $H \mathbin{/\mkern-6mu/} ab$. Also noted is that the magnetization is essentially linear with magnetic field for both $H \mathbin{/\mkern-6mu/} ab$ and $H \mathbin{/\mkern-6mu/} c$ without any signature of field-induced magnetic modifications (Figure S3).

Figure 2(b) presents the anisotropic electrical resistivity with current flow along the *ab* plane ($\rho_{ab}$) and *c* axis ($\rho_c$), respectively. Both $\rho_{ab}(T)$ and $\rho_c(T)$ show a similar temperature dependence, a bad-metal behavior with a broad hump at around 100 K. The anisotropic ratio $\rho_c/\rho_{ab}$ is about 30 above $T_N$, reflecting the Q2D electronic states. Below $T_N$, the anisotropy increases rapidly with decreasing temperature, indicating an enhanced 2D character in the AFM state. The AFM transition at $T_N$ is signaled by a rapid decrease in $\rho_{ab}$, which is more clearly visible in the plot of $d\rho_{ab}/dT$. The low-temperature resistivity obeys the power-law relation, $\rho = \rho_0 + A'T^\alpha$, with the fitted power $\alpha$ = 1.1 and 1.3 for $\rho_{ab}(T)$ and $\rho_c(T)$, respectively, suggestive of an unusual NFL behavior in the AFM state.

The Hall measurement data are presented in Figures 2(c) and 2(d). The Hall resistivity is almost linear with the magnetic field applied. Therefore, the Hall coefficient ($R_H$) can be well defined, and obtained by a linear fitting. The resulted $R_H(T)$ shows strong temperature dependence with a clear turning point (minimum) at $T_N$. This suggests existence of multiple Fermi-surface sheets (FSS) that are significantly influenced by the AFM ordering, consistent with an itinerant spin-density-wave scenario. The sign change in $R_H$ at about 20 K further confirms multiple FSS in the AFM state of $Cs_3V_9Te_{13}$.

We also measured the anisotropic magnetoresistance (MR) for the $Cs_3V_9Te_{13}$ crystals. As shown in Figure 2(e), for $H /\!/ c$, MR is positive at high temperatures, primarily due to additional carrier scattering by the external field. Below 130 K, MR becomes negative albeit of a small absolute value. This suggests the role of magnetism, which cancels out the aforementioned positive MR. In the case of $H /\!/ ab$ [see Figure 2(f)], MR is negligibly small at high temperatures, since the field-induced scattering is almost absent for the Q2D conduction. Negative MR develops below $T_N$, and the absolute MR values are almost one order of magnitude larger than those of $H /\!/ c$. The result further suggests in-plane magnetism in $Cs_3V_9Te_{13}$.

Figure 3(a) presents the specific-heat measurement data. An obvious anomaly is seen at $T_N$, confirming long-range magnetic ordering. Nevertheless, the released magnetic entropy up to $T_N$ appears to be small ($S_m \approx$ 1.38 J mol-fu$^{-1}$ K$^{-1}$, see Figure S2). Since the magnetic ordering happens most probably in the V2 sublattice, the expected magnetic entropy is $6R\ln(2S + 1) \approx 34.6$ J mol-fu$^{-1}$ K$^{-1}$ for the smallest local spin of $S = 1/2$. The greatly reduced value of $S_m$ can be explained by small-moment itinerant magnetism as well as existence of short-range magnetic correlations above $T_N$.

To extract the electronic specific-heat coefficient, i.e., the Sommerfeld coefficient, we plot $C/T$ as a function of $T^2$ in Figure 3(b). One sees a virtually linear relation except for the data below 2.8 K (the upturn below 2.8 K could be due to a nuclear Schottky anomaly). The linear fitting yields a Sommerfeld coefficient of $\gamma = 246$ mJ mol-fu$^{-1}$ K$^{-2}$, equivalent to 82 mJ (mol-CsV$_3$Te$_{4.33}$)$^{-1}$ K$^{-2}$. It is about four times of that of $CsV_3Sb_5$,[30] and close to that of $CsCr_3Sb_5$,[4] which highly suggests significant electron correlations in $Cs_3V_9Te_{13}$.

To further assess the strength of electron correlations in $Cs_3V_9Te_{13}$, we employ the concept of the Kadowaki-Woods (KW) ratio $A/\gamma^2$,[29,31,32] where $A$ is related to the carrier's scattering rate (obtained by fitting the low-temperature resistivity with $\rho = \rho_0 + AT^2$). For a rational comparison with $Ni_3In$ and $CsCr_3Sb_5$, the $\gamma$ value of 82 mJ (mol-CsV$_3$Te$_{4.33}$)$^{-1}$ K$^{-2}$ is taken. As shown in Figure 3(c), $CsV_3Te_{4.33}$ locates on the strong correlation regime represented by heavy-fermion compounds and correlated oxides.

As an intermetallic com-pound, notably, $CsV_3Te_{4.33}$ is very close to $Ni_3In$[29] and $CsCr_3Sb_5$,[4] both of which are kagome materials hosting FBs near $E_F$.

## Pressure Effect

We investigated the pressure effect in $Cs_3V_9Te_{13}$ to tune the strength of electron correlations and to explore possible superconductivity as well. Figure 4(a) shows the temperature dependence of resistivity under pressures applied by using piston-type cell and cubic anvil cell (CAC) respectively. With increasing pressure, the bad-metal behavior, characterized by the hump at around 100 K, is suppressed gradually. Concurrently, the AFM transition temperature $T_N$ moves to lower temperatures until it cannot be identified at 2.5 GPa. No superconductivity was observed down to 2 K and up to 9 GPa.

The low-temperature resistivity data (from 2 to 10 K) were fitted using the power-law formula, $\rho(T) = \rho_0 + A'T^\alpha$. The extracted parameters, $A'$ and $\alpha$, are plotted as functions of pressure in Figure 4(b). The $\alpha$ value is initially 1.2 at ambient pressure, as mentioned above. With increasing pressure, interestingly, $\alpha$ rapidly decreases to a minimum of 0.75 at $P_{c1}$ = 0.38 GPa. Concurrently, the coefficient $A'$ exhibits a maximum at $P_{c1}$. The result implies a quantum critical point (QCP) at $P_{c1}$, possibly arising from spin fluctuations of V1 atoms. The $\alpha$ value then increases gradually to ~1.8 at 2 GPa, followed by a small dip at 2.5 GPa. It further goes to above 2 for $P \geq 3.5$ GPa, accompanying with the decrease in $A'$. Note that, at high pressures when the magnetism is fully suppressed, the ubiquitous electron-phonon scattering comes to play a role, which could systematically raise the $\alpha$ values beyond the FL value of 2.0. Thus, we conjecture that the $\alpha$ value (1.6) at 2.5 GPa still an indication of NFL. There should be another QCP at $P_{c2}$ ~2.5 GPa, where $T_N$ is suppressed to zero. At higher pressures, $A'$ decreases continually, suggesting weakened electron correlations.

The above results are summarized in the electronic $P-T$ phase diagram shown in Figure 4(c). The evolution of the power $\alpha$ can be traced in the contour plot. Successive NFL-to-FL crossovers are seen in the low temperatures. Two possible QCPs appear inside the NFL states at $P_{c1} \approx 0.38$ GPa and $P_{c2} \approx 2.5$ GPa, respectively, probably associated with V1 and V2 sublattices. The absence of superconductivity might be due to the mutual interference between electrons of V1 and V2.

## DFT Calculations

To understand the experimental results, we performed density functional theory (DFT) calculations for $Cs_3V_9Te_{13}$. In the light of the relatively large unit cell containing two V sublattices with the dominant AFM interactions, it is not easy to find the magnetic ground state with DFT calculations. In this circumstance, below we presents the DFT calculation result for the nonmagnetic state of Cs$_3$V$_9$Te$_{13}$.

Figures 5(a)-(c) show the band structure and the density of states (DOS) of nonmagnetic Cs$_3$V$_9$Te$_{13}$, highlighting the contributions from 3$d$ orbitals of the V1 and V2 atoms as well as from each constituent element. First of all, the nearly flat bands along Γ−A indicate pronounced Q2D character, which accounts for the large anisotropy in resistivity. Secondly, the electronic states near $E_F$ are mostly contributed from vanadium 3$d$ electrons. To be specific, the $d_{yz/zx}$ and $d_{xy/x^2-y^2}$ electrons of V2 atoms dominate, while the 3$d_{z^2}$ electrons of V1 atoms contribute the remaining minor weight. Unexpectedly, two sets of kagome-like bands from the V2 sublattice can be recognized, as labelled in Figure 5(b). One of the topological FBs is just located at $E_F$. The total DOS at $E_F$ is $D(E_F)$ = 35.86 states eV$^{-1}$ fu$^{-1}$, corresponding to an electronic specific-heat coefficient of $\gamma_0$ = (1/3)$\pi k^2 D(E_F)$ = 84.6 mJ mol-fu$^{-1}$ K$^{-2}$. Thus the electron mass renormalization factor is $\gamma/\gamma_0 \approx 2.9$, indicating significant electron correlations in Cs$_3$V$_9$Te$_{13}$.

The appearance of kagome-like bands in Cs$_3$V$_9$Te$_{13}$ remind us of the previous theoretical study that shows two sets of kagome bands in a diatomic kagome lattice.[17,33,34] The diatomic kagome lattice is derived from a regular kagome lattice by splitting each lattice point into two, as shown in Figure 5(d). The V2 sublattice in Cs$_3$V$_9$Te$_{13}$ forms a twisted bipartite kagome lattice, in which neighboring V2 triangles rotate 42.4° clockwise and anticlockwise, respectively, as shown in Figure 5(e). Note that the resulted band structures only rely on the related interatomic hoppings. As a rough approximation, only $t_1$ and $t_2$ are considered here (both values should be quite similar because of the similar V2-V2 distances). According to the literature,[33] the schematic band structure for $t_2/t_1 \approx 0.75$ is displayed in Figure 5(f), which basically catch the main features of the DFT calculations. There are deviations from the theoretical model, which could arise from several factors including anisotropic hoppings of the $d$ orbitals, hybridizations with the V1 sublattice, long-range hoppings, and interlayer coupling.

To understand the pressure effect, we also calculated the band structure at 10 GPa. As shown in Figures (g) and (i), first of all, the DOS of V1 at $E_F$ becomes negligibly small, suggesting full delocalization of V1 electrons. Secondly, although the Dirac cone and vHs remain in Figure 5(h), the topological FBs from V2 are seriously broken, likely due to enhanced V1-V2 hybridizations and long-range hoppings. As a result, the $D(E_F)$ value [24.36 states eV$^{-1}$ fu$^{-1}$, equivalent to 8.12 states eV$^{-1}$ (CsV$_3$Te$_{4.33}$)$^{-1}$] is even smaller than that of CsV$_3$Sb$_5$,[20] consistent with the Pauli-paramagnetic state (Fermi liquid) observed experimentally.

# Summary

In summary, we have discovered a new vanadium-based material $Cs_3V_9Te_{13}$ which exhibits a cascade of correlated electron behaviors associated with the topological FB. The key structural unit is the $V_9Te_3$ plane which contains two interpenetrating sets of vanadium triangles. The V2 atoms reorganize into a twisted bipartite kagome lattice, generating two sets of kagome-like electronic bands. The Fermi level just locates at one of the FBs, giving rise to the magnetism and enhanced electron correlations. At ambient pressure, the material turns out to be a Q2D AFM bad metal with NFL behavior at low temperature and together with an enhanced Sommerfeld coefficient. Under hydrostatic pressure, the magnetism is sup-pressed, and quantum critical behaviors emerge with two possible QCPs, demonstrating high tunability in $Cs_3V_9Te_{13}$. The absence of superconductivity at around the QCPs provides in-sights into the conditions required for the emergence of unconventional superconductivity in correlated magnetically ordered systems. Future studies are required to clarify the magnetic order using neutron diffraction and nuclear magnetic resonance, directly probe the topo-logical FB using angle-resolved photoemission spectroscopy, and explore superconductivity using chemical doping and other tuning methods.

# Experimental and computation methods

Crystal growth was performed using a self-flux method with CsTe. Stoichiometric quantities of high-purity Cs (99.999%), V (99.9%), and Te (99.99%) elements in a ratio of 42 : 9 : 52 ratio were homogenized and loaded into an alumina crucible. The crucible was encapsulated in a Ta tube, which was weld-sealed in argon atmosphere to prevent oxidation. Then the Ta tube was enclosed in an evacuated quartz ampule. Thermal treatment involved heating to 1000 °C (20 h hold) followed by controlled cooling to 600 °C at 2 °C/h. Finally, the flux-grown crystals were isolated by washing with ion-free water. The obtained crystals showed no signs of degradation in air with size of up to $3 \times 3 \times 0.2$ mm$^3$ (Figure S1).

Chemical composition analysis was carried out via energy dispersive X-ray spectroscopy (EDS) using a Hitachi S-3700N scanning electron microscope equipped with an Oxford Instruments X-spectrometer. Structural characterization was performed by single-crystal X-ray diffraction (XRD) on a Bruker D8 Venture diffractometer utilizing Mo K$\alpha$ radiation at 220 K. Data reduction processes, which included integration and scaling, were carried out using APEX4 commercial software. To generate the (00$l$) diffraction pattern, $\theta - 2\theta$ scanning was conducted using a PANalytical X-ray diffractometer equipped with monochromatic Cu $K_{\alpha 1}$ radiation. Electrical resistivity, magnetoresistivity, Hall effect, and specific heat measurements were performed on a Quantum Design PPMS-9T system. Resistivity was measured via a standard four-terminal technique using graphite conductive adhesive for electrodes. Magnetic properties were characterized using a Quantum Design MPMS-3 system.

High-pressure resistivity with standard four-probe method were measured using commercial piston (< 2 GPa) and CAC (> 2 GPa) setup in SECUF.[35] In piston-type high-pressure cell, daphne 7373 was used as the pressure transmitting medium and superconducting transition temperature of Pb was measured to estimate the pressure. The pressure values in CAC were estimated from the pressure-loading force calibration curve at room temperature.

The DFT-based first-principles calculations were performed using Vienna ab initio Simulation Package.[36] The Kohn-Sham wave functions were treated with projected augmented wave method.[37] The exchange-correlation energy was calculated with a Perdew-Burke-Ernzerhof (PBE) type functional.[38] The energy cutoff of plane-wave basis was up to 600 eV and a Γ-centered $6 \times 6 \times 8$ k-point mesh was employed in the self-consistent calculations. The experimental room-temperature crystal structure was adopted for the ambient-pressure calculations. As for the high-pressure calculations, the lattice constants and atomic coordinates were fully relaxed with a solid revised PBE (PBEsol) functional,[39] leading to the calculated structural parameters of $a = 9.6600$ Å and $c = 7.3498$ Å.

## Acknowledgements

This work was supported by the National Key Research and Development Program of China (2022YFA1403202 and 2023YFA1406101) and the National Science Foundation of China (Grant No. 12474132). Part of the high-pressure measurements were carried out at the Cubic Anvil Cell station of Synergic Extreme Condition User Facility (SECUF, https://cstr.cn/31123.02.SECUF).## References

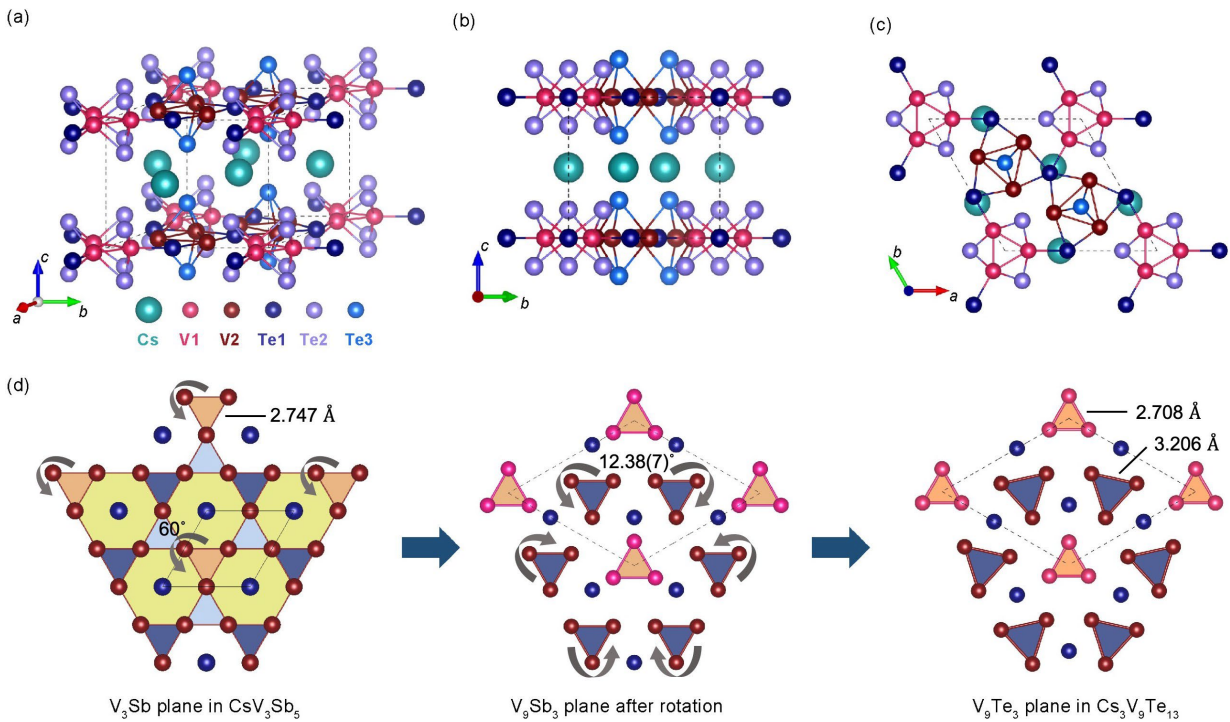

Figure 1: Crystal structure of $Cs_3V_9Te_{13}$. (a)-(c) display the views of 3D, $bc$, and $ab$ planes, respectively. The bottom panel (d) depicts the structure of $V_9Te_3$ planes at $z = 0$, which can be linked with an ideal kagome network in the prototype $CsV_3Sb_5$.

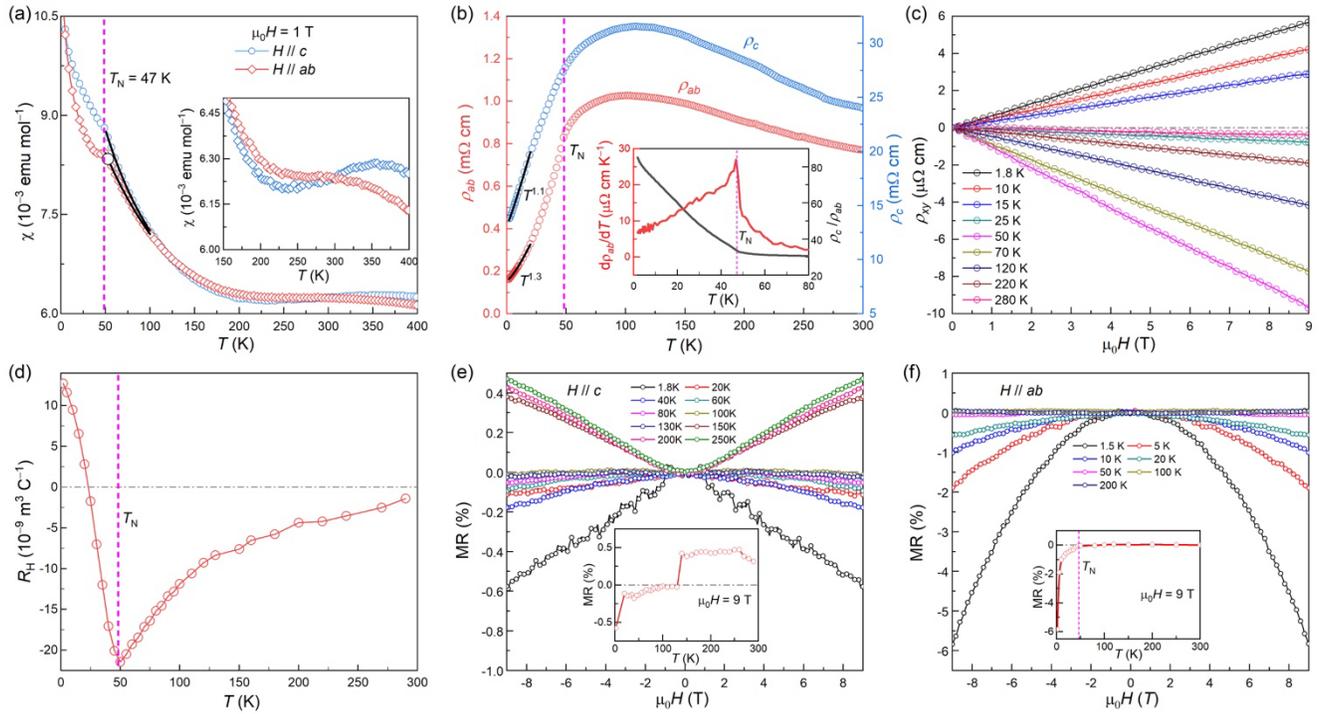

Figure 2: Magnetic susceptibility and electrical transport properties of $Cs_3V_9Te_{13}$ crystals. (a) Temperature dependence of the anisotropic susceptibility. The black solid lines represent Curie-Weiss fits using data from 50 to 140 K. The inset shows the close-up in the high-temperature region. (b) Temperature dependence of the anisotropic electrical resistivity. The solid lines in the low-temperature regime are power-law fits. The inset plots the derivative of $\rho_{ab}$ and the anisotropic ratio $\rho_c/\rho_{ab}$, from which $T_N = 47$ K can be identified (vertical dashed lines). (c) Field dependence of the Hall resistivity. (d) Temperature dependence of the Hall coefficient. (e), (f) Field dependence of magnetoresistance, defined by MR% = $100[\rho(H) - \rho(0)]/\rho(0)$, for $H \// c$ and $H \// ab$, respectively. The insets show the temperature dependence of MR at $\mu_0H = 9$ T.

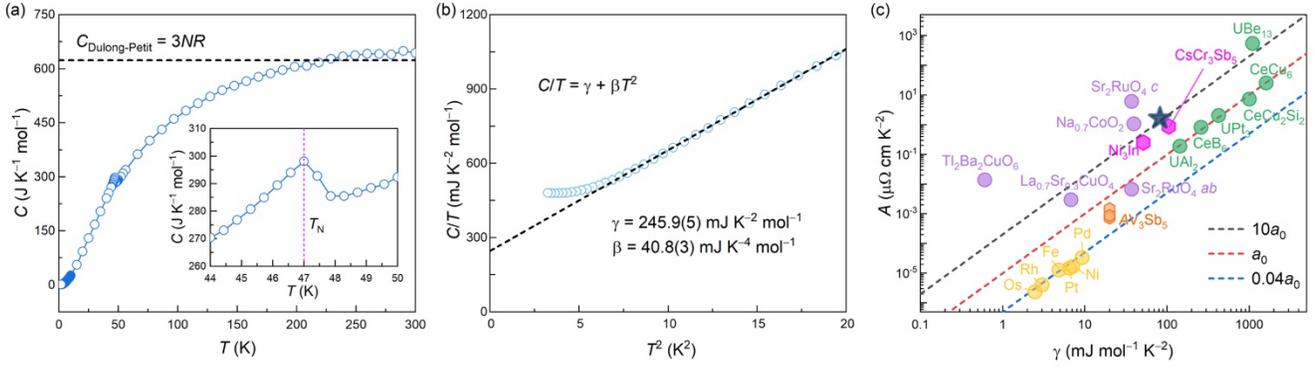

Figure 3: Specific-heat measurement of $Cs_3V_9Te_{13}$. (a) Temperature dependence of specific heat at zero field. The inset shows the close-up at around $T_N$. (b) Plot of $C/T$ versus $T^2$ in the low-temperature region. The dashed line is a linear fit from 2.8 to 4.5 K. (c) $A$-$\gamma$ diagram showing the Kadowaki-Woods ratio, $A/\gamma^2$, for various correlated materials. Part of data are adapted from Ref. [29] The data of $CsCr_3Sb_5$ are taken from Ref.[4] $a_0$ is a constant equal to 10 $\mu\Omega$ cm mol$^2$ K$^2$ J$^{-2}$.

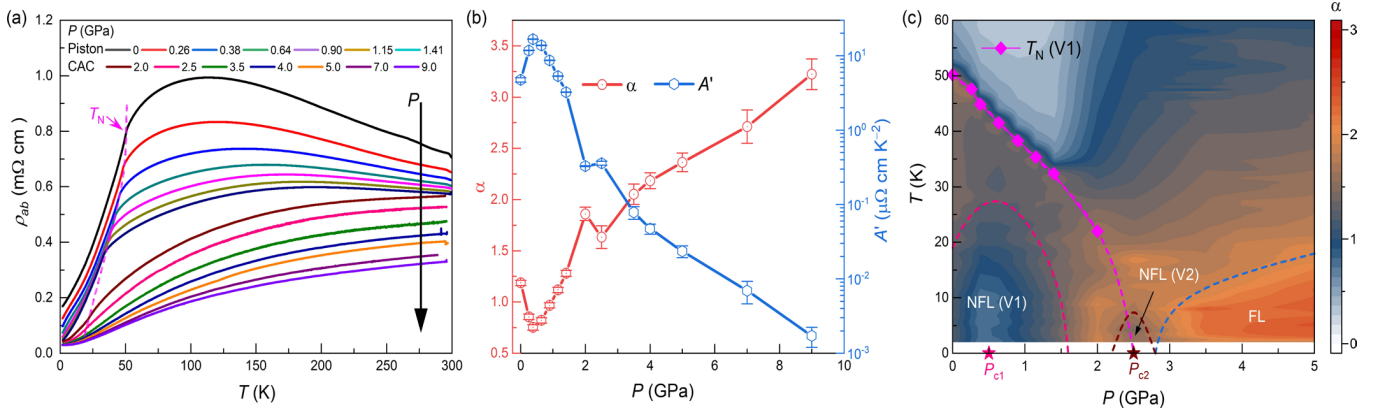

Figure 4: Pressure effect in $Cs_3V_9Te_{13}$. (a) Temperature dependence of in-plane resistivity under pressure. (b) Pressure dependence of $A'$ and $\alpha$, both of which are obtained from the data fitting (2 K $< T <$ 10 K) with the formula, $\rho = \rho_0 + A'T^\alpha$. (c) $P$-$T$ phase diagram. $\alpha$ is calculated by $d\ln(\rho - \rho_0)/d\ln T$, where $\rho_0$ is zero-temperature value via extrapolation. The black dashed lines are guides for the eye. NFL and FL stand for non-Fermi liquid and Fermi liquid, respectively.

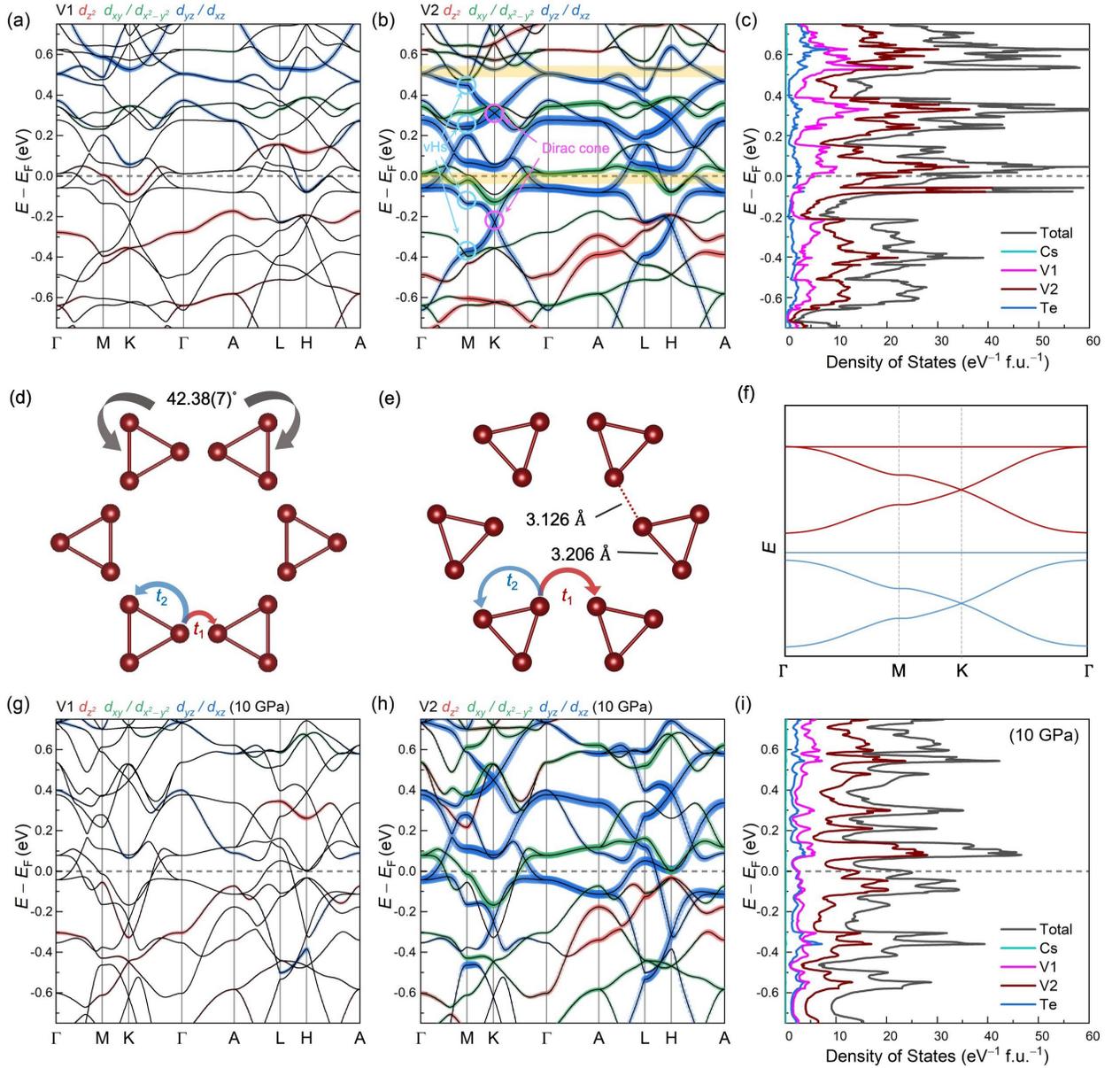

Figure 5: Electronic structures of $Cs_3V_9Te_{13}$ in relation with a bipartite kagome lattice. (a),(b) The original bipartite (diatomic) kagome lattice and its relation to the real structure of V2 triangles (V1 triangles are omitted). The resulted band structure ($t_2/t_1 \approx 0.75$, $t_3 = 0$) is schematically shown in (c).[26] (d)-(f) Calculated band structures and density of states of nonmagnetic $Cs_3V_9Te_{13}$ nearby $E_F$ highlighting the contributions from different $3d$ orbitals of V1 and V2. Panels (g)-(i) show the counterparts of (d)-(f) at 10 GPa.

# Supporting information

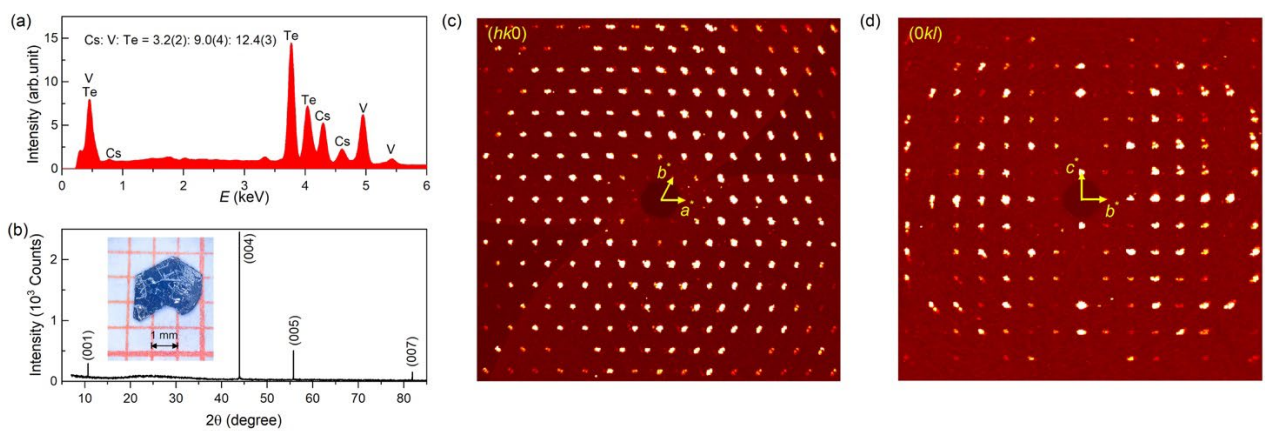

Figure S1: Characterizations of $Cs_3V_9Te_{13}$ by (a) energy-dispersive X-ray spectroscopy, (b) XRD ($00l$) reflections at room temperature (b), and reconstructed reciprocal space maps from single-crystal XRD at 220 K (c) and (d).

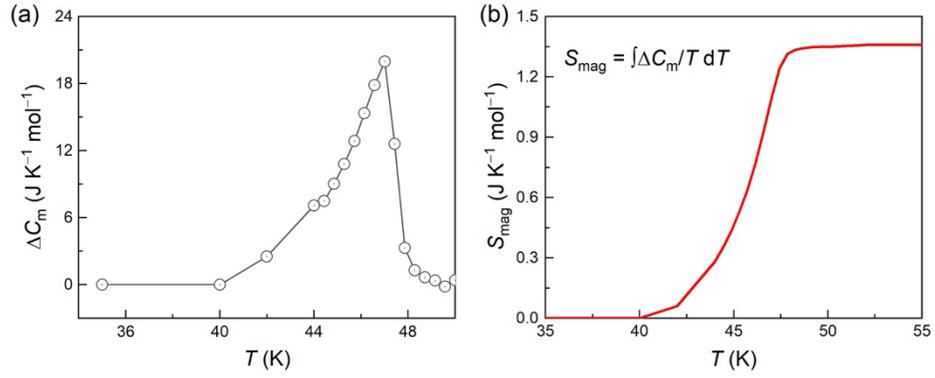

Figure S2: The magnetic entropy related to the AFM transition in $Cs_3V_9Te_{13}$. (a) Magnetic contribution of the specific heat by subtracting the phononic background. (b) Temperature dependence of released magnetic entropy across the magnetic transition.

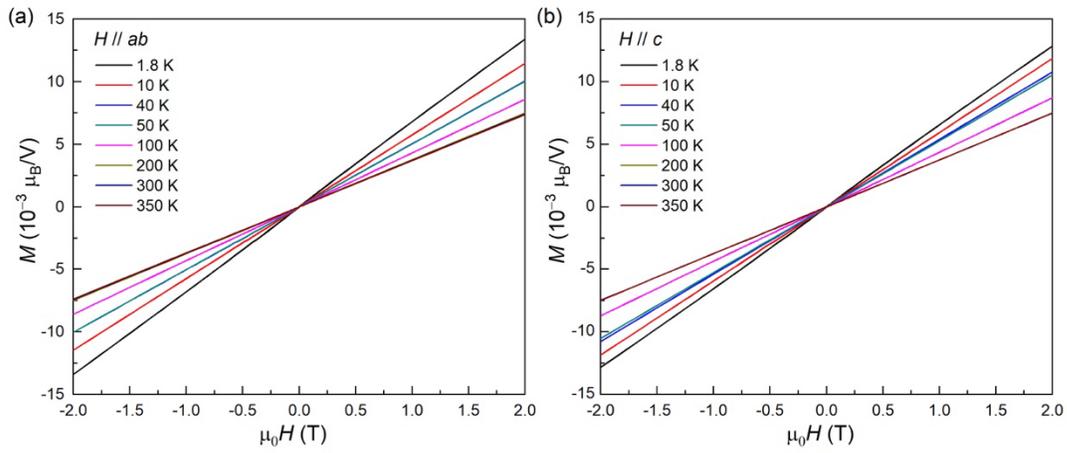

Figure S3: *M-H* curves from –2 T to 2 T at different temperatures with magnetic fields parallel to the *ab* plane (a) and *c* axis (b), respectively, for $Cs_3V_9Te_{13}$ crystals.

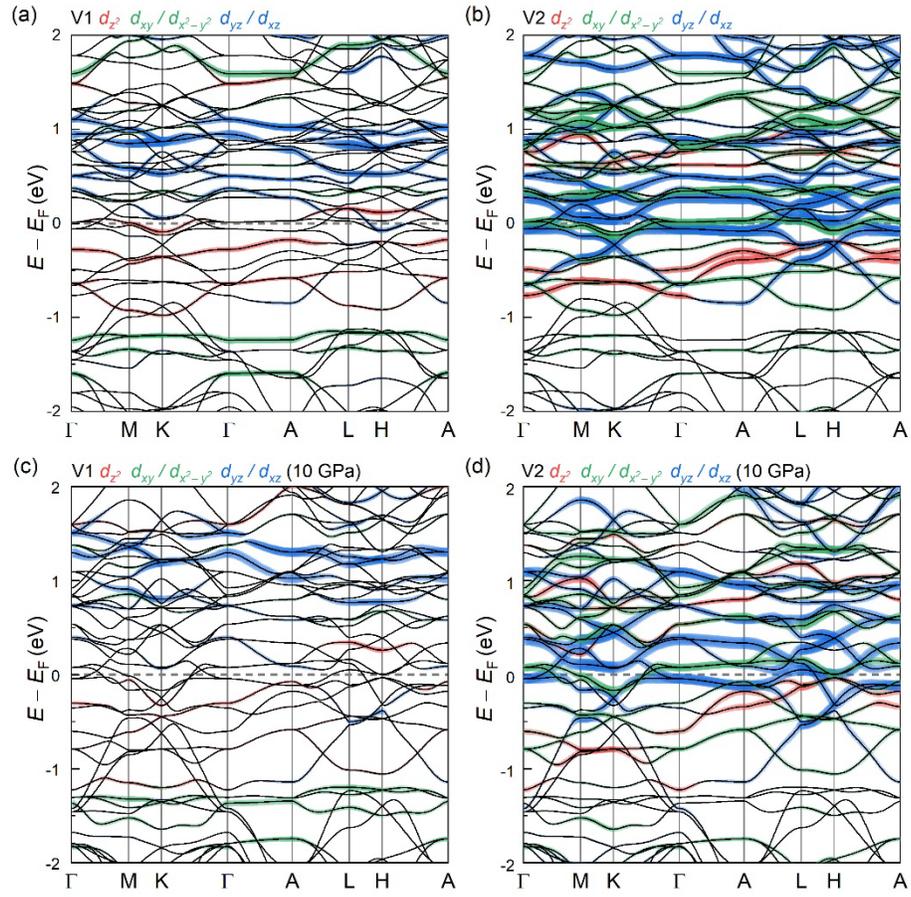

Figure S4: Calculated band structure of nonmagnetic $Cs_3V_9Te_{13}$ from −2 to 2 eV, highlighting contributions from V1 and V2 atoms at 0 GPa (top) and 10 GPa (bottom).

Table S1. Crystal data and structure refinement for $Cs_3V_9Te_{13}$ at 220 K.

| | |
|---|---|
| Empirical formula | $Cs_3V_9Te_{13}$ |
| Formula weight | 2515.98 |
| Temperature | 220 K |
| Wavelength | 0.71073 Å |
| Crystal system | $P\bar{6}2m$ |
| Space group | 189 |
| Unit cell dimensions | $a$ = 10.152(1) Å, $\alpha$ = 90° |
| | $b$ = 10.152(1) Å, $\beta$ = 90° |
| | $c$ = 8.204(3) Å, $\gamma$ = 120° |
| Volume | 732.2(4) Å$^3$ |
| Z | 1 |
| Density (calculated) | 5.706 g/cm3 |
| Absorption coefficient | 19.095 mm-1 |
| F(000) | 1048 |
| Crystal size | 0.164 x 0.106 x 0.015 mm3 |
| $\theta$ range for data collection | 2.317 to 30.531° |
| Index ranges | $-14 \leq h \leq 14, -14 \leq k \leq 14, -11 \leq l \leq 11$ |
| Reflections collected | 23036 |
| Independent reflections | 874 [$R_{int}$ = 0.0530] |
| Completeness to $\theta$ = 25.242° | 99.7% |
| Refinement method | F |
| Data / restraints / parameters | 874 / 0 / 29 |
| Goodness-of-fit | 1.119 |
| Final R indices [$I > 2\sigma(I)$] | $R_{obs}$ = 0.0234, $wR_{obs}$ = 0.0517 |
| R indices [all data] | $R_{all}$ = 0.0250, $wR_{all}$ = 0.0523 |
| Extinction coefficient | NA |
| Largest diff. peak and hole | 1.441 and -1.282 e·Å$^{-3}$ |

* $R = \Sigma||F_o| - |F_c||/\Sigma|F_o|, wR = \Sigma[w(|F_o|^2 - |F_c|^2)^2]/\Sigma[w(|F_o|^4)]^{1/2}$ and $w = 1/(\sigma^2(F) + 0.0001F^2)$.

| Label | x | y | z | Occupancy | $U_{eq}$* | $U_{11}$ | $U_{22}$ | $U_{33}$ | $U_{12}$ | $U_{13}$ | $U_{23}$ |
|---|---|---|---|---|---|---|---|---|---|---|---|
| Cs1 | 0 | 0.3631(2) | 0 | 1 | 29(1) | 20(1) | 32(1) | 30(1) | 10(1) | 0 | 0 |
| V1 | 0 | 0.1540(3) | 0.5 | 1 | 15(1) | 9(2) | 9(1) | 25(2) | 5(1) | 0 | 0 |
| V2 | 0.2882(2) | 0.4660(2) | 0.5 | 1 | 14(1) | 10(1) | 10(1) | 21(1) | 4(1) | 0 | 0 |
| Te1 | 0 | 0.4102(2) | 0.5 | 1 | 41(1) | 8(1) | 8(1) | 108(2) | 4(1) | 0 | 0 |
| Te2 | 0.2284(1) | 0.2284(1) | 0.7204(1) | 1 | 15(1) | 13(1) | 13(1) | 20(1) | 7(1) | 0 | 0 |
| Te3 | 1/3 | 2/3 | 0.2515(2) | 1 | 17(1) | 16(1) | 16(1) | 19(1) | 8(1) | 0 | 0 |

* $U_{eq}$ is defined as one third of the trace of the orthogonalized $U_{ij}$ tensor. The anisotropic displacement factor exponent takes the form: $-2\pi^2[h^2a^{*2}U_{11} + \ldots + 2hka^*b^*U_{12}]$.

| Bond | Distance (Å) | Bond | Distance (Å) | Bond | Distance (Å) |
|---|---|---|---|---|---|
| Cs1-Cs1 | 5.618(1) × 2 | V1-Te2 | 2.732(1) × 2 | V2-Te2 | 2.827(2) × 2 |
| Cs1-Te1 | 4.130(1) × 2 | V1-V2 | 3.053(2) × 2 | V2-Te3 | 2.754(2) × 3 |
| Cs1-Te2 | 3.959(1) × 3 | V1-V1 | 2.708(1) × 2 | V2-V2 | 3.206(4) × 6 |
| Cs1- Te3 | 3,844(1) × 4 | V2-Te1 | 2.666(2) × 2 | V2-V2 | 3.126(6) × 1 |
| V1-Te1 | 2.601(3) × 1 | V2-Te1 | 2.687(2) × 2 | | |

Table S2: Curie-Weiss fittings of susceptibility within different temperature range.

| | Temperature Range (K) | 50 – 100 | 50 – 120 | 50 – 140 | 50 – 150 | 50 – 160 |
|---|---|---|---|---|---|---|
| **H // ab** | $\chi_0$ ($10^{-3}$ emu mol$^{-1}$) | 3.0 | 3.6 | 4.2 | 4.2 | 4.5 |
| | $\mu_{\text{eff}}$ ($\mu_B$/V2) | 1.11 | 0.95 | 0.87 | 0.82 | 0.79 |
| | $\theta$ (K) | −120.40 | −94.18 | −79.21 | −70.89 | −63.78 |
| **H // c** | $\chi_0$ ($10^{-3}$ emu mol$^{-1}$) | 3.9 | 3.9 | 4.2 | 4.2 | 4.5 |
| | $\mu_{\text{eff}}$ ($\mu_B$/V2) | 0.87 | 0.82 | 0.78 | 0.76 | 0.74 |
| | $\theta$ (K) | −63.20 | −55.84 | −48.76 | −45.42 | −42.06 |